\documentclass[aps,prd,12pt,notitlepage,showpacs,nofootinbib,tightenlines]{revtex4-1}
\usepackage{amsmath}
\usepackage{amssymb}
\usepackage{epsfig}
\usepackage{graphicx}
\usepackage{bm}
\usepackage{times}
\usepackage{braket}
\usepackage{color}
\usepackage{slashed}
\usepackage{hyperref}
\definecolor{Red}{rgb}{1.,0.,0.}

\definecolor{Blue}{rgb}{0.,0.,1.}

\definecolor{nicered}{rgb}{0.7,0.1,0.1}
\definecolor{nicegreen}{rgb}{0.1,0.5,0.1}
\hypersetup{colorlinks,citecolor=nicegreen,linkcolor=nicered}

\begin{document}

\newcommand{\beq}{\begin{eqnarray}}
\newcommand{\eeq}{\end{eqnarray}}
\newcommand{\non}{\nonumber\\ }

\newcommand{\jpsi}{J/\Psi}

\newcommand{\ppa}{\phi_\pi^{\rm A}}
\newcommand{\ppp}{\phi_\pi^{\rm P}}
\newcommand{\ppt}{\phi_\pi^{\rm T}}
\newcommand{\ov}{ \overline }

\newcommand{\zerot}{ {\textbf 0_{\rm T}} }
\newcommand{\kt}{k_{\rm T} }
\newcommand{\kta}{{\textbf k_{\rm 1T}} }
\newcommand{\ktb}{{\textbf k_{\rm 2T}} }
\newcommand{\ktc}{{\textbf k_{\rm 3T}} }
\newcommand{\fb}{f_{\rm B} }
\newcommand{\fk}{f_{\rm K} }
\newcommand{\rk}{r_{\rm K} }
\newcommand{\mb}{m_{\rm B} }
\newcommand{\mw}{m_{\rm W} }
\newcommand{\im}{{\rm Im} }

\newcommand{\kks}{K^{(*)}}
\newcommand{\acp}{{\cal A}_{\rm CP}}
\newcommand{\pb}{\phi_{\rm B}}

\newcommand{\xeba}{\bar{x}_2}
\newcommand{\xsba}{\bar{x}_3}
\newcommand{\peas}{\phi^A}

\newcommand{\pvsl}{ p \hspace{-2.0truemm}/_{K^*} }
\newcommand{\esl}{ \epsilon \hspace{-2.1truemm}/ }
\newcommand{\Psl}{ P \hspace{-2truemm}/ }
\newcommand{\ksl}{ k \hspace{-2.2truemm}/ }
\newcommand{\lsl}{ l \hspace{-2.2truemm}/ }
\newcommand{\nsl}{ n \hspace{-2.2truemm}/ }
\newcommand{\vsl}{ v \hspace{-2.2truemm}/ }
\newcommand{\epsl}{\epsilon \hspace{-1.8truemm}/\,  }
\newcommand{\bfkk}{{\bf k} }
\newcommand{\calm}{ {\cal M} }
\newcommand{\calh}{ {\cal H} }

\def \appb{{\bf Acta. Phys. Polon. B }  }
\def \cpc{ {\bf Chin. Phys. C } }
\def \ctp{ {\bf Commun. Theor. Phys. } }
\def \epjc{{\bf Eur. Phys. J. C} }
\def \jhep{{\bf J. High Energy Phys. } }
\def \jpg{ {\bf J. Phys. G} }
\def \mpla{{\bf Mod. Phys. Lett. A } }
\def \npb{ {\bf Nucl. Phys. B} }
\def \npbps{ {\bf Nucl.Phys.B(Proc. Suppl.)} }
\def \npa{ {\bf Nucl. Phys. A} }
\def \plb{ {\bf Phys. Lett. B} }
\def \pr{  {\bf Phys. Rep.} }
\def \prc{ {\bf Phys. Rev. C }}
\def \prd{ {\bf Phys. Rev. D} }
\def \prl{ {\bf Phys. Rev. Lett.}  }
\def \ptp{ {\bf Prog. Theor. Phys. }}
\def \zpc{ {\bf Z. Phys. C}  }
\def \csb{ {\bf Chin. Sci. Bull. }}

\title{ The $\rho \gamma^* \to \pi (\rho)$ transition form factors in the Perturbative QCD factorization approach}
\author{Ya-Lan Zhang$^{1}$} 
\author{Shan Cheng$^{2}$} 
\author{Jun Hua$^{1}$} 
\author{Zhen-Jun Xiao$^{1,3}$ } \email{xiaozhenjun@njnu.edu.cn}
\affiliation{1.  Department of Physics and Institute of Theoretical Physics,
Nanjing Normal University, Nanjing, Jiangsu 210023, People's Republic of China,}
\affiliation{2.  Theoretische Elementarteilchenphysik, Naturwissenschaftlich Techn. Fakult$\ddot{a}$t,
Universi$\ddot{a}$t Siegen, 57068 Siegen, Germany,}
\affiliation{3. Jiangsu Key Laboratory for Numerical Simulation of Large Scale Complex Systems,
Nanjing Normal University, Nanjing 210023, People's Republic of China}
\date{\today}
\vspace{1cm}
\begin{abstract}
In this paper, we studied the $\rho \gamma^* \to \pi$ and $\rho\gamma^*\to \rho$
transition processes  and made the calculations for the  $\rho\pi$ transition form factor
$Q^4 F_{\rho\pi}(Q^2)$ and the $\rho$ meson electromagnetic form factors,
$F_{\rm LL, LT,TT}(Q^2)$ and $F_{1,2,3}(Q^2)$, by employing the perturbative QCD (PQCD) factorization
approach. For the $\rho \gamma^* \to \pi$ transition, we found that the contribution to form factor $Q^4 F_{\rho\pi}(Q)$
from the term proportional to the distribution amplitude combination $\phi^T_{\rho}(x_1)\phi^P_{\pi}(x_2)$
is absolutely dominant, and the PQCD predictions for both the size and the $Q^2$-dependence of this
form factor $Q^4 F_{\rho\pi}(Q^2)$ agree well with those from the extended ADS/QCD models or the light-cone QCD sum rule.
For the $\rho \gamma^* \to \rho$ transition and in the region of $Q^2\geq 3$ GeV$^2$, further more,
we found that the PQCD predictions for the magnitude and their $Q^2$-dependence of the $F_1(Q^2)$ and $F_2(Q^2)$
form factors agree well with those from the QCD sum rule, while the PQCD prediction for $F_3(Q^2)$
is much larger than the one from the QCD sum rule.
\end{abstract}



\maketitle

\section{Introduction}

During the past years, due to its very important role in understanding the hadron structure,
the various hadron form factors have been studied intensively for example in
Refs.~\cite{plb72-368,npb216-373,npa618-291}.
The transition and the electromagnetic form factors of the pseudoscalar mesons, especially the
pion meson as the lightest QCD bound state, attracted the most attentions theoretically
\cite{plb094-245,prd60-074004,prd70-033014,prl65-1717,plb115-410,plb328-457,
9909450,prd79-034015,prd72-054506,prd74-034008,npb821-291,prd83-054029,jhep-1401}
and experimentally\cite{prd9-1229,prl97-192001,prl95-261803}.
The transition form factors between the pseudoscalar and vector mesons are also
investigated by employing rather different approaches
\cite{zpc20-357,jpg34-1845,epjc6-477,epjc24-117,epjc67-253,prd75-094020},
the resulted theoretical predictions however are self-consistent and comparable from each other.
The radiative form factors of vector meson, such as the $\rho$ meson, also draw some interests
\cite{plb420-8,prd70-033001}.

The $k_T$ factorization theorem\cite{kt-theorem,prd67-034001} is one of the major factorization
approaches based on the factorization hypothesis\cite{prd22-2157,pr112-173,prl43-545}
and the resummation image in the end-point region\cite{prd47-3875,prd66-094010}.
Due to it's clear advantages, such as no end-point singularity and can provide a large
strong phase to generate the sizable CP violation for B meson decays, the PQCD factorization approach
based on the $k_T$ factorization theorem has been widely used to study the two-body
hadronic decays of $B/B_s/B_C$ mesons for example in 
Refs.~\cite{prd52-5358,prd67-094013,plb712-63,prd65-014007, prd63-074009,prd70-093013,xiao2006,bs,liux,prd89-074046}.
Recently, the next-to-leading-order (NLO) corrections to some important hadron form factors
have been calculated ~\cite{prd83-054029,prd85-074004,plb718-1351,npb896-255}.
With the inclusion of these NLO corrections, the PQCD predictions for the heavy-to-light pseudoscalar
form factors in $B \to \pi$ transition, for example, become well consistent with those from the
light-cone QCD sum rule \cite{zpc60-349,prd71-014015,prd83-094031,jhep04-014}.
In this paper, we will study the $\rho \pi$ transition from factor and the $\rho$ meson
electromagnetic form factor by employing the pQCD factorization approach.

This paper is organized as follows. The relevant kinetics and the meson wave functions
are introduced in Sec.~II. In Sec.~III, the pseudoscalar and vector transition form factors
corresponding to the process $\rho \gamma^{\star} \to \pi$ are studied with the
use of the two-parton meson wave functions up to sub-leading twist.
In Sec.~IV, the $\rho$ meson electromagnetic form factors in $\rho \gamma^{\star} \to \rho$
process are calculated. Sec.~V contains the summary of this paper.

\section{Kinetics and input wave functions}
\begin{figure}[tb]
\vspace{-2cm}
\begin{center}
\leftline{\epsfxsize=16cm\epsffile{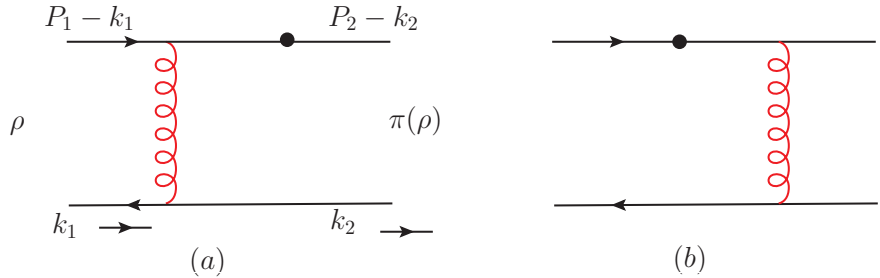}}
\end{center}
\vspace{-18cm}
\caption{The LO diagrams for the $\rho\pi$ transition- and $\rho$ meson electromagnetic form factor,
with the symbol $\bullet$ representing the virtual photon vertex.}
\label{fig:fig1}
\end{figure}
In this section, we first consider the relevant kinetics and the input meson wave functions to be used
in our calculation.
The leading-order(LO) topological diagrams for the meson transition or electromagnetic form factors are
presented in Fig.~\ref{fig:fig1}.
Of course, the fully diagrams should also include the other two sub-diagrams with the photon vertex being on
the lower anti-quark lines.
Because of the isospin symmetry, these extra two sub-diagrams have same structure as Fig.~\ref{fig:fig1}(a,b)
with exchanging quark and anti-quark in the initial and final mesons, the only difference may be appeared is the quark charge.

Under the light-cone(LC) coordinate, the momentum for initial and final mesons in Fig.~\ref{fig:fig1} are defined as,
\beq
P_1&=& \frac{Q}{\sqrt{2}}(1,r^2_{\pi/\rho},\mathbf{0}_T), \qquad P_2=\frac{Q}{\sqrt{2}}(r^2_{\pi/\rho},1,\mathbf{0}_T);\non
k_1&=&(x_1\frac{Q}{\sqrt{2}},0,\mathbf{k}_{1T}), \qquad k_2=(0,x_2\frac{Q}{\sqrt{2}},\mathbf{k}_{2T}),
\label{eq:kinetic-momentum}
\eeq
where $k_{i},i=1,2$ is the momentum carried by the anti-quark for initial or final meson with the momentum fraction $x_i$,
and $\mathbf{k}_{iT}$ represent the corresponding transversal momentum.
The dimensionless parameter $\gamma_{\pi}^2 \equiv M_{\pi}^2/Q^2, \gamma_{\rho}^2 \equiv M_{\rho}^2/Q^2$.
Then the momentum transfer squared is $q^2=(P_1-P_2)^2=-Q^2$.
The polarization vectors of the initial and final  $\rho$ mesons are also defined in the LC coordinate:
\beq
\epsilon_{1\mu}(L)=\frac{1}{\sqrt{2}\gamma_{\rho}}(1,-\gamma_{\rho}^2,\mathbf{0}_{T}),~~~~~\epsilon_{1\mu}(T)=(0,0,\mathbf{1}_{T});\non
\epsilon_{2\mu}(L)=\frac{1}{\sqrt{2}\gamma_{\rho}}(-\gamma_{\rho}^2,1,\mathbf{0}_{T}),~~~~~\epsilon_{2\mu}(T)=(0,0,\mathbf{1}_{T}).
\label{eq:kinetic-polarization}
\eeq
with $\rho$ meson mass $M_{\rho}=0.77$ GeV.
The LC definitions in Eq.~(\ref{eq:kinetic-momentum},\ref{eq:kinetic-polarization}) satisfy the relations,
\beq
P_{1} \cdot \epsilon_{1}=0,~~~~~P_{2} \cdot \epsilon_{2}=0;~~~~~\epsilon^2_{1}=\epsilon^2_{2}=-1.
\label{eq:kinetic-relations}
\eeq

As elaborated in Ref.~\cite{plb712-63}, the power-suppressed three-parton contribution to the pion electromagnetic form factor
in the $k_T$ factorization theorem can only provide $\sim 5\%$ correction to the LO form factor
in the whole range of experimentally accessible momentum transfer squared.
This sub-leading piece amount only up to few percents of the $B \to \pi$ transition form factor at large recoil of
the pion in $k_T$ factorization theorem.
So in our calculation for the $\rho\pi$ transition form factor and $\rho$ meson electromagnetic form factors,
we can just consider the dominant contributions from the two-parton meson wave functions
and neglect the very small three-parton part safely.
The initial wave functions for transversal and longitudinal polarized $\rho$ meson can be written as,
\beq
\Phi_{\rho}(P_{1},\epsilon_{1}(L))&=&\frac{1}{\sqrt{6}}
\Bigl [ \epsl_{1}(L) M_{\rho} \phi_{\rho}(x_1) +  \epsl_{1}(L) \Psl_{1} \phi^{t}_{\rho}(x_1) +
                                                          M_{\rho} \phi^{s}_{\rho}(x_1) \Bigr ],
\label{eq:rho-l1} \\
\Phi_{\rho}(P_{1},\epsilon_{1}(T))&=&\frac{1}{\sqrt{6}}\Bigl [ \epsl_{1}(T) \Psl_{1} \phi^{T}_{\rho}(x_1) + \epsl_{1}(T) M_{\rho} \phi^{v}_{\rho}(x_1) \non
&&  \hspace{1cm}+ i M_{\rho} \epsilon_{\mu\nu\rho\sigma} \gamma^{\mu} \gamma_5 \epsilon^{\nu}_{1}(T) n^{\rho} v^{\sigma} \phi^{a}_{\rho}(x_1)
\Bigr ],
\label{eq:irho-t}
\eeq
The final pion or rho meson wave function are written as,
\beq
\Phi_{\pi}(P_{1})&=&\frac{1}{\sqrt{6}}\left[ \gamma_5 \Psl_{2} \phi^{A}_{\pi}(x_2) + \gamma_5 m_0^{\pi} \phi^{P}_{\phi}(x_2) -
                                         \gamma_5 m_0^{\pi} (\vsl\nsl-1) \phi^{T}_{\pi}(x_2) \right];
\label{eq:final-pion-wf} \\
\Phi_{\rho}(P_{2},\epsilon_{2}(L))&=&\frac{1}{\sqrt{6}}\left[ \epsl_{2}(L) M_{\rho} \phi_{\rho}(x_2) +  \epsl_{2}(L) \Psl_{2} \phi^{t}_{\rho}(x_2) +
                                                          M_{\rho} \phi^{s}_{\rho}(x_2) \right],
\label{eq:rho-l2}\\
\Phi_{\rho}(P_{2},\epsilon_{2}(T))&=&\frac{1}{\sqrt{6}}\Bigl [ \epsl_{2}(T) \Psl_{2} \phi^{T}_{\rho}(x_2) + \epsl_{2}(T) M_{\rho} \phi^{v}_{\rho}(x_2)
\non
&& \hspace{1cm} + i M_{\rho} \epsilon_{\mu\nu\rho\sigma} \gamma_5 \gamma^{\mu} \epsilon^{\nu}_{2}(T)
v^{\rho} n^{\sigma} \phi^{a}_{\rho}(x_2) \Bigr ].
\label{eq:frho-t}
\eeq
Here $\phi^{A}_{\pi}$, $\phi^{T}_{\rho}$ and $\phi_{\rho}$ are the leading twist-2 (T2) distribution amplitudes(DAs),
while $\phi^{P,T}_{\pi}$, $\phi^{v,a}_{\rho}$ and $\phi^{t,s}_{\rho}$ are the sub-leading twist-3 (T3) DAs
which are power suppressed by $m_0^{\pi}/Q$ and $M_{\rho}/Q$. And $m_0^{\pi}$ is the the chiral mass of pion meson.

The pion meson DAs with the inclusion of the high order
effects as given in Ref.~\cite{jhep01-010} are adopted in our numerical calculation:
\beq
\phi_{\pi}^{A}(x) &= &\frac{3 f_{\pi} }{\sqrt{6}}
x (1-x) \left [ 1 +  a_2^{\pi} C_2^{\frac{3}{2}}(t) + a_4^{\pi} C_4^{\frac{3}{2}}(t) \right ], \non
\phi_{\pi}^{P}(x) &= &\frac{f_{\pi} }{2\sqrt{6} }
\left[ 1 + \left (30 \eta_3 - \frac{5}{2} \rho_{\pi}^2 \right ) C_2^{\frac{1}{2}}(t) -
3 \left (\eta_3 \omega_3 + \frac{9}{20} \rho_{\pi}^2 \left (1 + 6 a_2^{\pi}\right )
\right ) C_4^{\frac{1}{2}}(t) \right ],  \non
\phi_{\pi}^{T}(x) &= & \frac{f_{\pi} }{2\sqrt{6}} (1-2x) \left [1 + 6 \left (5 \eta_3 - \frac{1}{2} \eta_3 \omega_3 - \frac{7}{20} \rho_{\pi}^2
- \frac{3}{5} \rho_{\pi}^2 a_2^{\pi} \right ) \left (1-10 x + 10 x^2 \right )  \right ],
\label{eq:DAs-pion}
\eeq
where the new Gegenbauer moments $a_i^\pi$, the parameters $\eta_3, \omega_3$ and $\rho_\pi$ are adopted as\cite{jhep04-014}:
\beq
a_2^{\pi} &=& 0.16, \quad a_4^{\pi} = 0.04, \quad
\rho_{\pi} = m_{\pi}/m_0, \quad \eta_3 = 0.015, \quad \omega_3 = -3.0,
\label{eq:input-pion}
\eeq
with $f_\pi=0.13$ GeV, $m_\pi=0.13$ GeV,  $m^0_\pi=1.74$ GeV.

For the rho meson, the following twist-2 DAs ( $\phi_\rho$ and $\phi_\rho^T$ ) and twist-3
DAs ( $\phi_\rho^{v,a,t,s}$) will be used in our numerical calculation:
\beq
\phi_{\rho}(x) &=& \frac{3f_{\rho}}{\sqrt{6}} x(1-x) [1+a^{\|}_{2\rho}C^{3/2}_{2}(t)],\non
\phi^{T}_{\rho}(x)&=& \frac{3f^{T}_{\rho}}{\sqrt{6}} x(1-x) [1+a^{\bot}_{2\rho}C^{3/2}_{2}(t)],
\label{eq:rho-t2}\\
\phi^{v}_{\rho}(x)&=&\frac{f_{\rho}}{2\sqrt{6}} [0.75(1+t^{2}) + 0.24(3t^2-1) + 0.12(3-30t^2+35t^4)],\non
\phi^{a}_{\rho}(x)&=&\frac{4f_{\rho}}{4\sqrt{6}} (1-2x) [1+0.93(10x^2-10x+1)];\non
\phi^{t}_{\rho}(x)&=& \frac{f^{T}_{\rho}}{2\sqrt{6}} \left [ 3t^{2} + 0.3t^2(5t^2-3) + 0.21(3-30t^2+35t^4) \right ],\non
\phi^{s}_{\rho}(x)&=& \frac{3f^{T}_{\rho}}{2\sqrt{6}}(1-2x) \left [1 + 0.76\left ( 10x^2-10x+1 \right )\right ],
\label{eq:rho-t3}
\eeq
where $t=2x-1$, the Gegenbauer moments $a^{\|}_{2\rho}=0.18, a^{\bot}_{2\rho}=0.2$ and the decay constants
$f_{\rho}=0.209, f_{\rho}^T=0.165$. The Gegenbauer polynomials in Eqs.~(\ref{eq:rho-t2},\ref{eq:rho-t3}) are of the following
form:
\beq
C^{1/2}_2(t)&=&\frac{1}{2}[3t^2-1], \quad C^{3/2}_2(t)=\frac{3}{2}[5t^2-1],\non
C^{1/2}_4(t)&=&\frac{1}{8}[3-30t^2+35t^4], \quad C^{3/2}_4(t)=\frac{15}{8}[1-14t^2+21t^4].
\label{eq:ci01}
\eeq

\section{The $\rho\pi$ transition form factor $Q^4 F_{\rho\pi}(Q)$ }

We firstly consider the $\rho \gamma^* \to \pi$ transition, here the two sub-diagrams Fig.~\ref{fig:fig1}(a,b)
will contribute to the $\rho\pi$ transition form factor.
The final state is a pseudoscalar pion meson, which can't be generated by a scalar current,
so only the transversal polarized initial $\rho$ meson with the vector current $J_{\mu, |\lambda|=1}$ contributes to
$\rho \pi$ transition form factor, which can be written as,
\beq
<\pi(P_2)|J_{\mu}|\rho(P_1,\epsilon_1)>&=&<\pi(P_2)|J_{\mu, |\lambda|=1}|\rho(P_1,\epsilon_1(T))> \non
&=&F_{\rho\pi}(Q^2)\epsilon_{\mu\nu\rho\sigma}\epsilon^{\nu}_{1}(T)n^{\rho}v^{\sigma} P^+_1 P^-_2.
\label{eq:rho-pion-ff-definition}
\eeq
For the case of the large momentum transfer, the asymptotic behavior of the hadron form factors is the
form of \cite{pr112-173}
\beq
<P_2,\lambda_2|J_{\lambda}(0)|P_1,\lambda_1> \sim \left(\frac{1}{\sqrt{|q^2|}}\right)^{|\lambda_1-\lambda_2|+2n-3}.
\label{eq:asymptotic-behaviour-ff}
\eeq
The $\rho\pi$ transition amplitude is suppressed by $1/Q^2$ because of the helicity flipping at the vector vertex
for the quark lines ($|\lambda|=1$).
In Eq.~(\ref{eq:asymptotic-behaviour-ff}), $\lambda_1$ and $\lambda_2$ denote the helicity on the z-axis
and n is the parton number of hadron: when the hadron is a meson, $n=2$.

From Eqs.~(\ref{eq:rho-pion-ff-definition},\ref{eq:asymptotic-behaviour-ff}) one can see that the $\rho\pi$
transition form factor $F_{\rho\pi}(Q^2)$
has the asymptotic behaviour $[Q]^{-4}$ at the limit of large transfer momenta,
so one should study the dimensionless form factor $Q^4 F_{\rho\pi}(Q^2)$ rather than $F_{\rho\pi}(Q^2)$ itself.
After the inclusion of the contributions from all sub-diagrams, Fig.~(\ref{fig:fig1})(a,b) and their partner diagrams
with the vertexes on the lower anti-quark lines,  the vector and pseudoscalar $\rho\pi$ transition hard kernel can be
written in the following form:
\beq
Q^4H_{\rho\pi}(Q;x_1,x_2;\mathbf{k}_{1T},\mathbf{k}_{2T})&=&\frac{16\pi \alpha_s }{3} \Bigl \{
                     \frac{M_{\rho} [\phi^{v}_{\rho}(x_1)-\phi^a_{\rho}(x_1)] \phi^A_{\pi}(x_1)}{(P_1-k_2)^2(k_1-k_2)^2} \non
&& \hspace{-1cm} + \frac{x_1M_{\rho}[\phi^v_{\rho}(x_1)-\phi^a_{\rho}(x_1)]\phi^A_{\pi}(x_2) +
                           2m^0_{\pi}\phi^T_{\rho}(x_1)\phi^P_{\pi}(x_2)}{(P_2-k_1)^2(k_1-k_2)^2}   \Bigr\}.
\label{eq:rho2pi-ff}
\eeq

By integrating over the longitudinal momentum fractions $(x_1,x_2)$ and the transversal momentum in
it's conjugate coordinate spaces $(b_1,b_2)$, we can obtain the $\rho\pi$ transition form factor:
\beq
Q^4 F_{\rho\pi}(Q) &=& \frac{16 \pi }{3} Q^4 \cdot \alpha_s(\mu) \cdot \int^1_0 dx_1 dx_2 \int^{\infty}_0 b_1 db_1 b_2 db_2
                      \cdot \textmd{exp}[-S(x_i;b_i;Q;\mu)] \non
&& \times \Bigl \{ M_{\rho} \bigl[ \phi^{v}_{\rho}(x_1)-\phi^a_{\rho}(x_1) \bigr]\cdot
                  \phi^A_{\pi}(x_1) \cdot h(x_2,x_1,b_2,b_1) \non
&& +  x_1M_{\rho}\left[ \phi^v_{\rho}(x_1)-\phi^a_{\rho}(x_1) \right]\cdot \phi^A_{\pi}(x_2) \cdot  h(x_1,x_2,b_1,b_2)\non
&& + 2m_0^{\pi}\phi^T_{\rho}(x_1)\phi^P_{\pi}(x_2)\cdot S_{t}(x_1)\; S_{t}(x_2) \cdot  h(x_1,x_2,b_1,b_2)
                    \Bigr\},
\label{eq:rho2pi-ff1}
\eeq
where the $k_T$ resummation Sudakov factor $S(x_i;b_i;Q;\mu)$ and the threshold resummation function $S_t(x)$
are the same ones as being used in Refs.~\cite{prd65-014007,prd83-054029,npb896-255}.
In numerical calculation we choose $c=0.4$ in the function $S_t(x)$.
The hard function $h(x_1,x_2,b_1,b_2)$ in Eq.~(\ref{eq:rho2pi-ff1}) can be  written as the following form:
\beq
h(x_1,x_2,b_1,b_2)&=&K_0\left (\sqrt{x_1x_2} Q b_2 \right)
\Bigl [\theta(b_1-b_2)I_0\left (\sqrt{x_1} Q b_1 \right) K_0\left (\sqrt{x_1} Q b_2 \right) + (b_1 \leftrightarrow b_2)\Bigr],
\eeq
where the function $K_0$ and $I_0$ are the modified Bessel function.
Following Refs.~\cite{prd65-014007,prd83-054029,npb896-255},
we here also choose $\mu$ and $\mu_f$ as the largest hard scale in the numerical calculations:
\beq
\mu=\mu_f= t = \max\left (\sqrt{x_1} Q, \sqrt{x_2}Q, 1/b_1,1/b_2 \right).
\label{eq:mumuf}
\eeq

\begin{figure*}
\centering
\vspace{-0.5cm}
\includegraphics[width=0.5\textwidth]{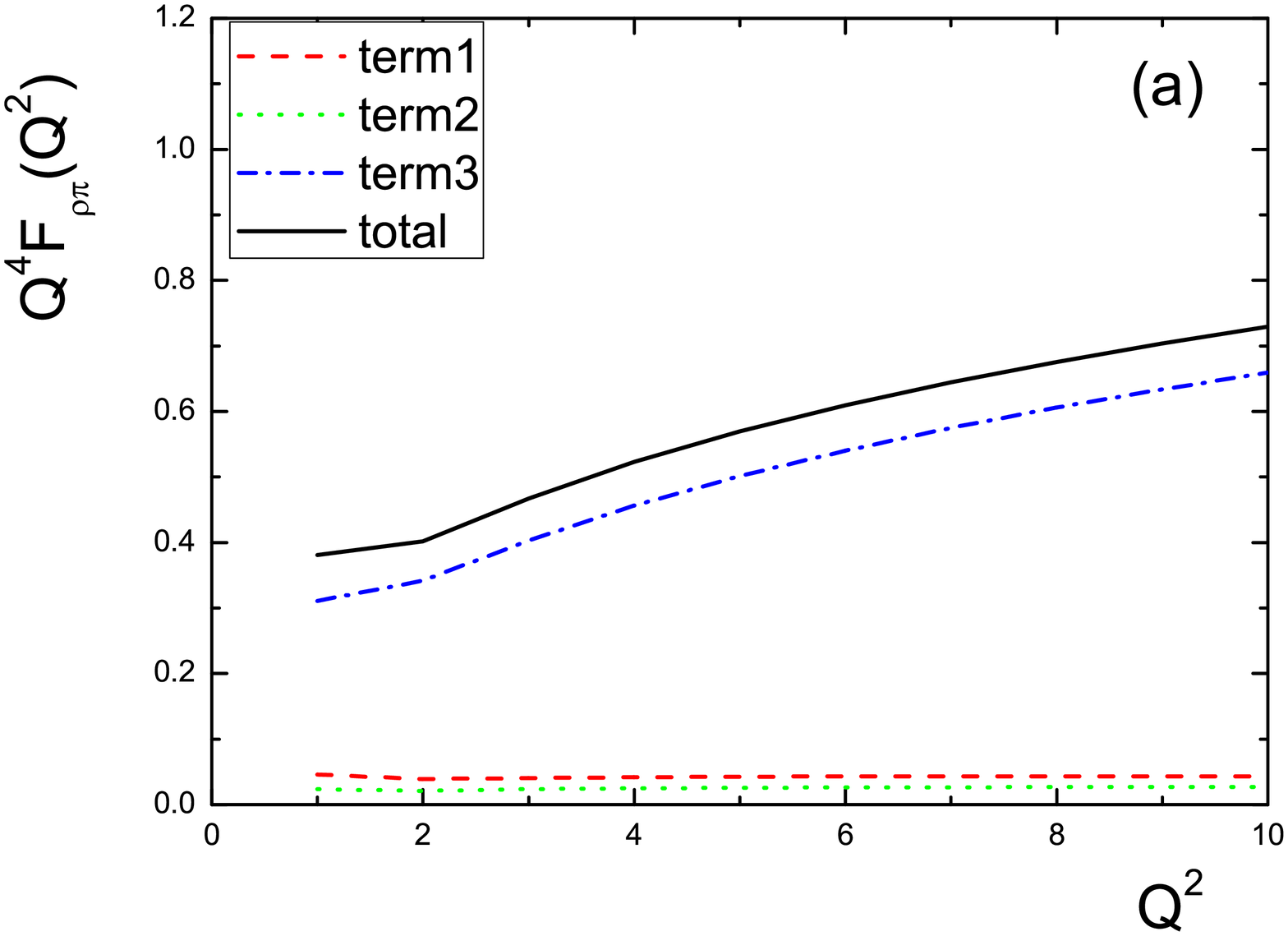}
\hspace{-0.5cm}
\includegraphics[width=0.5\textwidth]{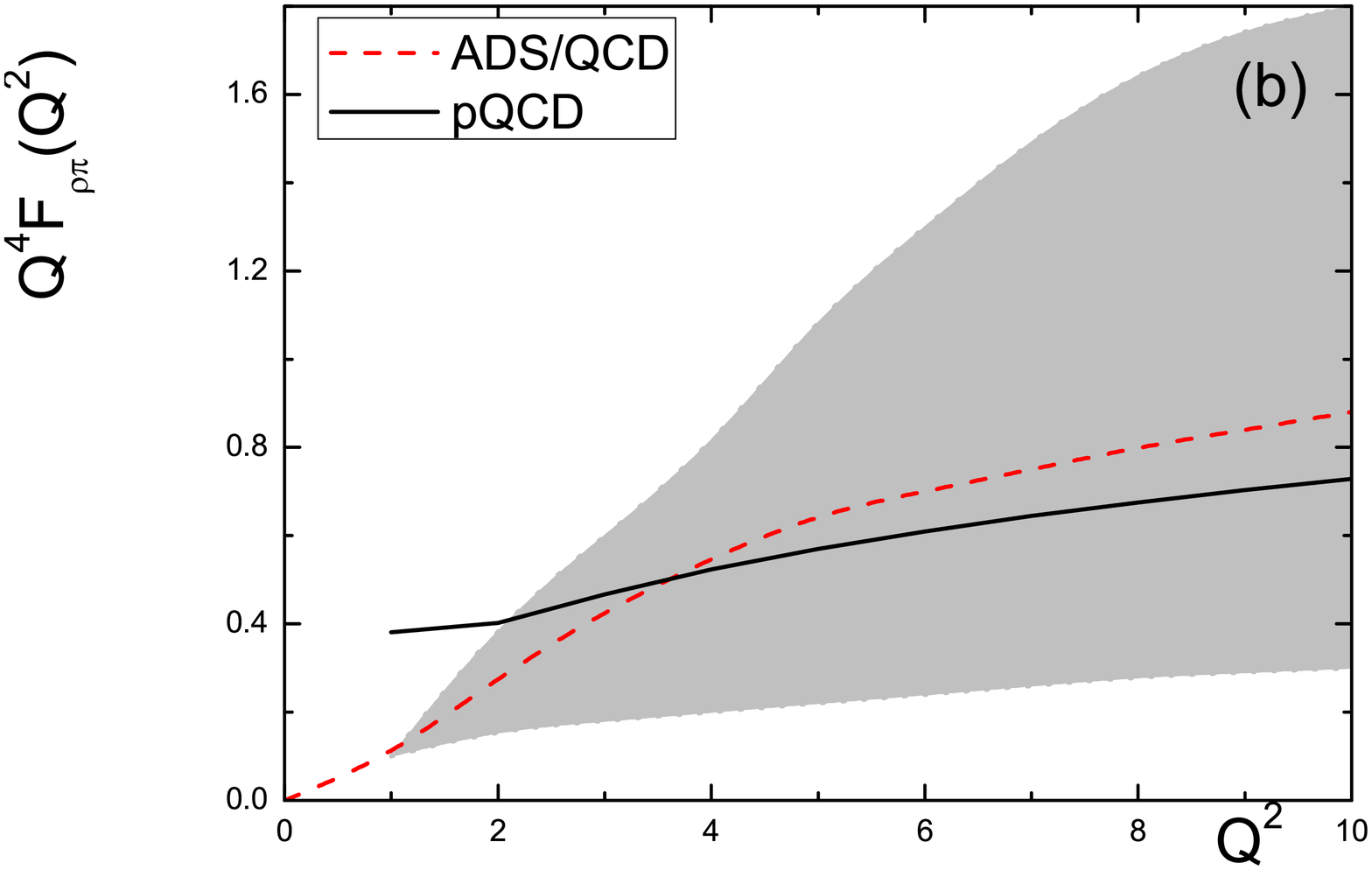}
\vspace{-0.5cm}
\caption{(a) the PQCD predictions for $Q^2$ dependence of the $\rho\pi$ transition form factor $Q^4F_{\rho\pi}(Q^2)$;
and (b) the theoretical predictions from ADS/QCD (dashed curve) or the ligh-cone QCD sum rules (dark region),
and from the PQCD factorization approach (solid curve).}
\label{fig:fig2}
\end{figure*}

Based on the formula in Eq.~(\ref{eq:rho2pi-ff1}), we calculate and show the PQCD predictions for the
$Q^2$-dependence of the $\rho\pi$ transition form factor $Q^4\; F_{\rho\pi}(Q)$ in Fig.~\ref{fig:fig2}.
In Fig.~\ref{fig:fig2}(a), the dashed-curve shows the contribution from the first term of Eq.~(\ref{eq:rho2pi-ff1}),
corresponding to the T3$\&$T2 product term from Fig.~\ref{fig:fig1}(a) and its partner with the virtual vertex being on the lower anti-quark line;
while the dot-dashed and dots curve shows the contribution from the second and the third term of Eq.~(\ref{eq:rho2pi-ff1}),
coming from the Fig.~\ref{fig:fig1}(b) and its partner. The solid curve in Fig.~\ref{fig:fig2}(a) refers to the total contribution.
In Fig.~\ref{fig:fig2}(b), the dashed curve shows the theoretical prediction based on the ADS/QCD theory \cite{epjc67-253},
while the dark-region shows the theoretical predictions from the light-cone QCD sum rules
\cite{epjc6-477,plb328-457,zpc20-357}. The PQCD prediction ( the solid curve in Fig.2(b)) is drawn here as a comparison.

From the curves in Fig.~\ref{fig:fig2}, one can see the following points:
\begin{enumerate}
\item[1]
The third term in Eq.~(\ref{eq:rho2pi-ff1}) with the DAs combination $\phi^T_{\rho}(x_1)\phi^P_{\pi}(x_2)$
provide the absolutely dominant contribution. The first term describe the contribution from Fig.~\ref{fig:fig1}(a) and
is very small in size.

\item[2]
The PQCD predictions for both the magnitude and the $Q^2$-dependence of the $\rho\pi$ transition form factor
$Q^4\; F_{\rho\pi}(Q)$ agree well with the theoretical predictions obtained in the extended ADS/QCD models
\cite{epjc67-253} or in the classical light-cone QCD sum rule \cite{epjc6-477,plb328-457,zpc20-357}.

\end{enumerate}

\section{$\rho$ meson electromagnetic form factor $F_i(Q^2)$}

In this section we consider the $\rho\gamma^* \to \rho$ process, and
calculate the $\rho$ meson electromagnetic form factors by employing the PQCD factorization approach.
Because the initial and final states are the same $\rho$ meson, we only investigate Fig.~\ref{fig:fig1}(a) in detail,
the contributions from the other three topological diagrams can be obtained from the exchanging symmetry
\cite{prd83-054029,npb896-255} from Fig.~\ref{fig:fig1}(a).
The lorentz invariant $\rho$ electromagnetic form factors $F_i(Q^2), i=1,2,3$ in $\rho\gamma^* \to\rho$ process
are defined as the following matrix element \cite{prd70-033001,npb216-373},
\beq
&&<\rho(P_2,\epsilon^{\star}_2)|J_{\mu,\lambda}|\rho(P_1,\epsilon_1)> = - \epsilon_{2\beta}\epsilon_{1\alpha} \Bigl\{
   \left [ (P_1+P_2)_{\mu}g^{\alpha\beta}-P_2^{\alpha}g^{\beta\mu}-P_1^{\beta}g^{\alpha\mu} \right ]\cdot F_1(Q^2) \non
&&\hspace{1.5cm}  - \left [ g^{\mu\alpha}P_1^{\beta}+g^{\mu\beta}P_2^{\alpha} \right ] \cdot F_2(Q^2)
  +\frac{1}{M_{\rho}^2}P_2^{\alpha}P_1^{\beta}(P_1+P_2)_{\mu} \cdot F_3(Q^2) \Bigr \}.
\label{eq:rho-em}
\eeq

From the asymptotic behavior as defined in Eq.~(\ref{eq:asymptotic-behaviour-ff}), one can see that
the transversal form factors are suppressed by $1/Q^2$ or $1/Q^3$:
\beq
&&<\rho(P_2,\lambda_2=0)|J_{\mu,|\lambda|=1}|\rho(P_1,|\lambda_1|=1)> \sim \frac{1}{Q^2},
\label{eq:ts-1}\\
&&<\rho(P_2,\lambda_2=\pm 1)|J_{\mu,|\lambda|=0}|\rho(P_1,\lambda_1=\mp 1)> \sim \frac{1}{Q^3},
\label{eq:ts-2}
\eeq
while the asymptotic behavior of the longitudinal form factor is the normal one:
\beq
<\rho(P_2,\lambda_2=0)|J_{\mu,|\lambda|=0}|\rho(P_1,\lambda_1=0)> \sim \frac{1}{Q}.
\label{eq:l-normal}
\eeq
The helicity of the vector current is defined as $\lambda=\lambda_1+\lambda_2$,
then the transition $\lambda_1=\pm 1 \to \lambda_2=\pm 1$ is forbidden because the helicity of vector current is $\lambda \leq 1$.
The $\lambda_1=\pm1 \to \lambda_2=0$ transition in Eq.~(\ref{eq:ts-1}) requires the helicity flipping
for one quark line at the vector vertex, which give a suppression $k_T/Q$;
While the $\lambda_1=\pm 1 \to \lambda_2= \mp 1$ transition in Eq.~(\ref{eq:ts-2}) needs the helicity flipping
for both quark lines at the vector vertices, which leads to a suppression $k^2_T/Q^2$.

From the radiative matrix element as defined in Eq.~(\ref{eq:rho-em}) and the asymptotic behavior of the hadron form factors
as described in Eqs.~(\ref{eq:ts-1},\ref{eq:ts-2},\ref{eq:l-normal}), we get to know that:
\begin{enumerate}
\item[(i)]
Only the electric form factor $F_1(Q^2)$ contributes to the transversal from factor $F_{\rm TT}(Q^2)$
with $\lambda_1=-\lambda_2=\pm 1$;

\item[(ii)]
Both the electric form factor $F_1(Q^2)$ and the magnetic form factor $F_2(Q^2)$ contribute to the semi-transversal form
factor $F_{\rm LT}$ with $\lambda_i=\pm 1 ~ \& ~ \lambda_j=0 (i,j=1,2, i \neq j)$;

\item[(iii)]
All $F_1(Q^2), F_2(Q^2)$ and the quadruple form factor $F_3(Q^2)$ give the contribution to the
longitudinal radiation form factor $F_{\rm LL}$ with $\lambda_1=\lambda_2=0$.
\end{enumerate}
These form factors do satisfy the following relations by definition:
\beq
F_{\rm TT}(Q^2)&=&F_1(Q^2),\non
F_{\rm LT}(Q^2)&=&\frac{Q}{2M_{\rho}} \left[ F_1(Q^2) + F_2(Q^2) \right],\non
F_{\rm LL}(Q^2)&=&F_1(Q^2) - \frac{Q^2}{2M^2_{\rho}}F_2(Q^2) + \frac{Q^2}{M^2_{\rho}}\left(1+\frac{Q^2}{4M^2_{\rho}}\right)F_3(Q^2).
\label{eq:ff-relations}
\eeq

By introducing the transversal momentum $k_T$ to cancel the end-point singularity, integrating over the longitudinal momentum fractions
and transversal coordinate conjugated to transversal momentum, one finds the PQCD predictions for the
$Q^2$ dependence of the $\rho$ meson  electromagnetic from factors:
\beq
F_{\rm LL}(Q^2)&=&\frac{32 \pi C_F}{3} Q^2 \alpha_s(\mu) \int^1_0 dx_1 dx_2 \int^{\infty}_0 b_1 db_1 b_2 db_2
                       ~\cdot \textmd{exp}[-S(x_i;b_i;Q;\mu)]\non
&~& \times \Bigl \{  \bigl [ x_2-\frac{1}{2}\gamma^2_{\rho}(1+x_2) \bigr ] \phi_{\rho}(x_1)\phi_{\rho}(x_2) + \gamma^2_{\rho}\phi^s_{\rho}(x_1)\phi^t_{\rho}(x_2) \non
&~&                 + 2\gamma^2_{\rho}(1-x_2)\phi^s_{\rho}(x_1)\phi^s_{\rho}(x_2)  \Bigr\} \cdot h(x_2,x_1,b_2,b_1) , \label{eq:rhoff-LL} \\
F_{\rm LT}(Q^2)&=&\frac{32 \pi C_F}{3} Q M_{\rho} \alpha_s(\mu) \int^1_0 dx_1 dx_2 \int^{\infty}_0 b_1 db_1 b_2 db_2
                       ~\cdot \textmd{exp}[-S(x_i;b_i;Q;\mu)] \non
&~& \times \Bigl \{  \frac{1}{2} \left[\phi^v_{\rho}(x_1)+\phi^a_{\rho}(x_1)\right] \phi_{\rho}(x_2)
                    +\phi^s_{\rho}(x_1)\phi^T_{\rho}(x_2) \non
&&        \ \ \     - \frac{1}{2}\phi_{\rho}(x_1) \left[ \phi^v_{\rho}(x_2)+\phi^a_{\rho}(x_2)\right] \Bigr\} \cdot h(x_2,x_1,b_2,b_1), \label{eq:rhoff-LT}\\
F_{\rm TT}(Q^2)&=&\frac{32 \pi C_F}{3} M^2_{\rho} \alpha_s(\mu) \int^1_0 dx_1 dx_2 \int^{\infty}_0 b_1 db_1 b_2 db_2
                       ~\cdot \textmd{exp}[-S(x_i;b_i;Q;\mu)] \non
&~& \times   \Bigl \{  (1-x_2) \left[\phi^a_{\rho}(x_1)\phi^a_{\rho}(x_2)-\phi^v_{\rho}(x_1)\phi^v_{\rho}(x_2)\right ] \non
&~&        \ \ \     + (1+x_2) \left[\phi^a_{\rho}(x_1)\phi^v_{\rho}(x_2)-\phi^v_{\rho}(x_1)\phi^a_{\rho}(x_2) \right ]
\Bigr\} \cdot h(x_2,x_1,b_2,b_1),
\label{eq:rhoff-TT}
\eeq
where $C_F=4/3$, the Sudakov factor $S(x_i;b_i;Q;\mu)$ and the hard function $h(x_2,x_1,b_2,b_3)$ are the same
one as those appeared in Eq.~(\ref{eq:rho2pi-ff1}). With these form factors of different polarized
initial and final states and the relations in Eq.~(\ref{eq:ff-relations}),
one can obtain easily the $Q^2$ dependence of Lorentz-invariant electric, magnetic and quadruple form factors
$F_1(Q^2), F_2(Q^2)$ and $F_3(Q^2)$.

\begin{figure*}
\centering
\vspace{-0.5cm}
\includegraphics[width=0.6\textwidth]{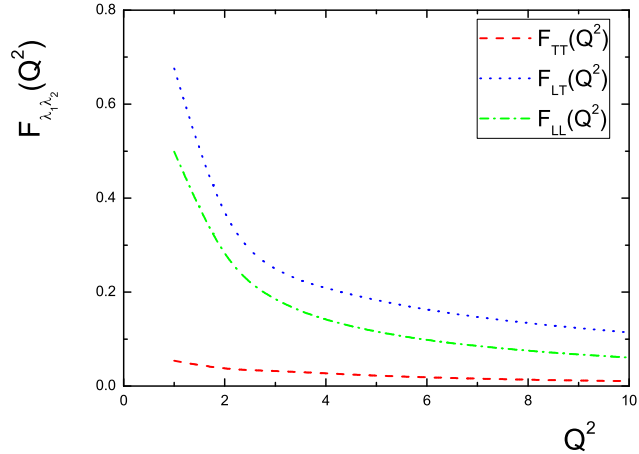}
\vspace{-0.5cm}
\caption{The PQCD predictions for $Q^2$ dependence of the $\rho$ meson form factors $F_{\lambda_1\lambda_2}(Q^2)$
with different polarizations for initial and final states. The short dashed, dots and dot-dashed curve
represents the form factor $F_{\rm TT}(Q^2)$,  $F_{\rm LT}(Q^2)$ and $F_{\rm LL}(Q^2)$, respectively.}
\label{fig:fig3}
\end{figure*}
\begin{figure*}
\centering
\vspace{-0.5cm}
\includegraphics[width=0.6\textwidth]{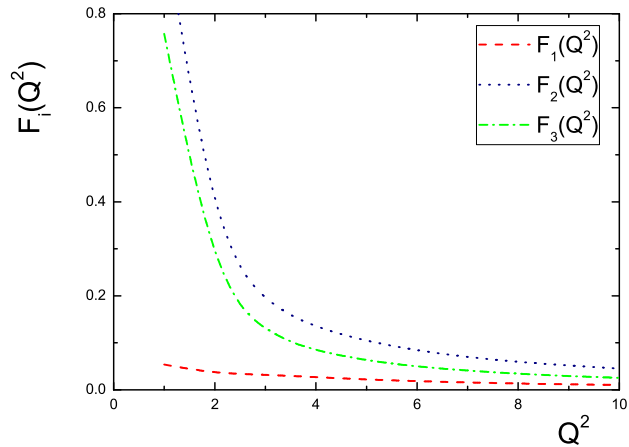}
\vspace{-0.5cm}
\caption{The PQCD predictions for $Q^2$ dependence of the Lorentz-invariant $\rho$ meson electromagnetic form
factors $F_{1,2,3}(Q^2)$. The short dashed, dots and dot-dashed curve
represents the electric ($F_1(Q^2)$), the magnetic ($F_2(Q^2)$) and the quadruple ($F_3(Q^2)$) form factor,
respectively.}
\label{fig:fig4}
\end{figure*}

The PQCD predictions for the $Q^2$-dependence of the $\rho$ meson electromagnetic form factors with different polarizations
(i.e. $F_{\rm LL}(Q^2)$,  $F_{\rm LT}(Q^2)$ and $F_{\rm LL}(Q^2)$ ) are presented in Fig.~\ref{fig:fig3}, while
the $Q^2$-dependence of the $\rho$ meson electric, magnetic and quadruple
form factors with different Lorentz structures ( i.e. $F_{1,2,3}(Q^2)$ ) are presented in Fig.~\ref{fig:fig4}.
From these two figures, one can find the following points:
\begin{enumerate}
\item [(i)]
For the form factors with different polarizations, there exists  an approximate relation:
$F_{\rm LT}(Q^2) \gtrsim F_{\rm LL}(Q^2) \gg F_{\rm TT}(Q^2)$.
The asymptotic behavior displayed in Eq.~(\ref{eq:asymptotic-behaviour-ff}) is partially violated for $F_{\rm LT}(Q^2)$ and
$F_{\rm LL}(Q^2)$.
This violation arose because the longitudinal form factor $F_{\rm LL}$ has an additional suppression from $x_2$,
although its asymptotic behavior has a $Q/M_{\rho}$ enhancement when  compared to the semi-transversal form factor $F_{\rm LT}$.
And this violation is consistent with the light-cone sum rule result \cite{npb216-373}.

\item [(ii)]
The value of the electric form factor $F_1(Q^2)$ is rather small, because it happens being equal to the
heavily suppressed transversal form factor $F_{\rm TT}(Q^2)$.
In the region of $Q^2 \geq 3 $ GeV$^2$, there exists  also an approximate relation for the values of $F_{1,2,3}(Q^2)$:
$F_2(Q^2) \gtrsim F_3(Q^2) \gtrsim F_1(Q^2)$.
Our PQCD predictions for $F_1(Q^2)$ and $F_2(Q^2)$ form factors agree well with the
QCD sum rule results \cite{prd70-033001} in the region $Q^2 \geq 3$ GeV$^2$.
But the PQCD prediction for the quadruple form factor $F_3(Q^2)$ is much larger than the one from the QCD sum rule,
the hierarchy between $F_1(Q^2)$ and $F_3(Q^2)$ predicted in the QCD sum rule is therefore turn over here.

\end{enumerate}

\section{Summary}

By employing the pQCD factorization approach, we studied the $\rho \gamma^* \to \pi$ and the $\rho \gamma^* \to \rho$
transition processes and made the analytical and numerical evaluations for the  $\rho\pi$ transition form factor
$Q^4 F_{\rho\pi}(Q^2)$ and the $\rho$ meson electromagnetic form factors, $F_{\rm LL, LT,TT}(Q^2)$ and $F_{1,2,3}(Q^2)$.
We found the following results:
\begin{enumerate}
\item[(i)]
For the $\rho \gamma^* \to \pi$ transition process, the contribution to the $\rho\pi$ transition form
factor $Q^4\; F_{\rho\pi}(Q)$ from the terms proportional to the DAs combination
$\phi^T_{\rho}(x_1)\phi^P_{\pi}(x_2)$ is absolutely dominant,
and the PQCD predictions for both the magnitude and the $Q^2$-dependence of this form factor agree well with those from
the extended ADS/QCD models or from the light-cone QCD sum rule.

\item[(ii)]
For the $\rho \gamma^* \to \rho$ transition process and in the region of $Q^2 \geq 3$GeV$^2$,
we found that the PQCD predictions for the magnitude and their $Q^2$-dependence of the $F_1(Q^2)$ and $F_2(Q^2)$ 
form factors agree well with those from the QCD sum rule, while the PQCD prediction for the quadruple $F_3(Q^2)$ 
is much larger than the one from the QCD sum rule.
\end{enumerate}

\section{Acknowledement}

The authors would like to thank Hsiang-nan Li and Cai-Dian L\"u for valuable discussions.
This work is supported by the National Natural Science Foundation of China under Grant No.11235005
and by the Project on Graduate Students¡¯ Education and Innovation of Jiangsu
Province under Grant No. KYZZ15-0212.


\end{document}